\documentclass[aps, prb, amsmath, twocolumn, floatfix, longbibliography, superscriptaddress, footinbib, 10pt]{revtex4-2}

\usepackage{ifthen}
\usepackage{amssymb, amsmath,amsfonts}
\usepackage[per-mode=symbol]{siunitx} 
\usepackage[final]{microtype} 
\usepackage{setspace}
\usepackage{xcolor}
\usepackage[utf8]{inputenc}
\usepackage[l2tabu, orthodox]{nag}
\usepackage{physics}
\usepackage{mathtools}
\usepackage{bm}
\usepackage{upgreek}
\usepackage{booktabs} 

\usepackage{array} 
\usepackage{adjustbox}
\usepackage[flushleft]{threeparttable}
\usepackage{threeparttablex} 
\usepackage{comment}
\usepackage{calc}

\usepackage{lmodern}      
\usepackage[T1]{fontenc}  

\usepackage{graphicx}
\usepackage{epstopdf}  

\usepackage{flafter,placeins}
\usepackage{float}
\usepackage{color,soulutf8,colortbl,xspace}
\usepackage[skins]{tcolorbox}
\usepackage{multirow}

\makeatletter
\let\ps@plain=\ps@pagenumber
\makeatother

\usepackage[acronym]{glossaries}
\usepackage{blindtext}
\usepackage[colorlinks=true,linkcolor=blue, citecolor=blue, urlcolor=blue]{hyperref}
 
\newcommand{\polydept}{Department of Engineering Physics, \'Ecole Polytechnique de Montr\'eal, C.P. 6079, Succ. Centre-Ville, Montr\'eal, Qu\'ebec, Canada H3C 3A7}

\newcommand{\GeSn}[2]{Ge$_{#1}$Sn$_{#2}$}
\newcommand{\quot}[1]{``#1''} 
\newcommand{\kp}{\textit{k.p}}

\newcommand{\comments}[1]{} 
\newcommand{\mentionSupp}[1]{\textcolor{blue}{Supporting information S#1}}

\newcolumntype{g}{>{\columncolor{lightgray}} c}

\makeatletter
\newcommand*{\DivideLengths}[2]{%
  \strip@pt\dimexpr\number\numexpr\number\dimexpr#1\relax*65536/\number\dimexpr#2\relax\relax sp\relax
}
\makeatother

\newcommand{\fref}[2]{\autoref{#1}\textcolor{blue}{#2}}

\newcommand{\paperSection}[2][normal]{
    \ifthenelse{\equal{#1}{normal}}{
        \medskip\noindent{\textbf{#2}}\newline
    }{
        \noindent\textbf{#2}\newline
    }
}

\newcommand{\citesupp}{
    \footnote{See Supplemental Material for more details about the momentum matrix elements and the special points approximation}
}

\newcommand{\ac}[1]{\gls*{#1}}

\graphicspath{{./figures}{./figures/downsized_image/}}

\hypersetup{
  pdftitle={Radiative Carrier Lifetime in GeSn Mid-Infrared Emitters},
  pdfauthor={Gérard Daligou, Anis Attiaoui, Simone Assali, Patrick Del Vecchio, Oussama Moutanabbir},
  pdfkeywords={Radiative carrier lifetime, Mid-infrared, Germanium tin semiconductors, k,p theory, Joint Density of States, Photoluminescence},
  bookmarksnumbered,
  pdfstartview={FitV},
  linktoc=all
}

\makeglossaries

\begin{document}

\newacronym{QW}{QW}{quantum wells}
\newacronym{SBEMA}{SBEMA}{single-band effective mass approximation}
\newacronym{MQW}{MQW}{multi Quantum Wells}
\newacronym{DH}{DH}{double heterostructure}
\newacronym{PBA}{PBA}{parabolic band approximation}
\newacronym{BZ}{BZ}{Brillouin zone}
\newacronym{SLA}{SLA}{special-lines approximation}
\newacronym{EL}{EL}{electroluminescence}
\newacronym{PL}{PL}{photoluminescence}
\newacronym{IVBA}{IVBA}{inter-valence bands absorption}
\newacronym{FD}{FD}{Fermi-Dirac distribution}
\newacronym{EFA}{EFA}{envelope function approximation}
\newacronym{JDOS}{JDOS}{joint density of states}
\newacronym{SRH}{SRH}{Shockley-Read-Hall}
\newacronym{TL}{TL}{top layer}
\newacronym{ML}{ML}{middle layer}
\newacronym{BL}{BL}{bottom layer}
\newacronym{CVD}{CVD}{chemical vapor deposition}
\newacronym{PVD}{PVD}{physical vapor deposition}
\newacronym{MBE}{MBE}{molecular beam epitaxy}
\newacronym{LP-CVD}{LP-CVD}{low-pressure chemical vapor deposition}
\newacronym{PECVD}{PECVD}{plasma-enhanced chemical vapor deposition}
\newacronym{UHV}{UHV}{ultra-high vacuum}
\newacronym{RSM}{RSM}{reciprocal space mapping}
\newacronym{XRD}{XRD}{X-ray diffraction}
\newacronym{STEM}{STEM}{scanning transmission electron microscope}
\newacronym{FWHM}{FWHM}{full width at half maximum}
\newacronym{TE}{TE}{transverse electric}
\newacronym{TM}{TM}{transverse magnetic}
\newacronym{FTIR}{FTIR}{Fourier Transform InfraRed}
\newacronym{MWIR}{MWIR}{mid-wave infrared}
\newacronym{NIR}{NIR}{near-infrared}
\newacronym{SWIR}{SWIR}{short-wave infrared}
\newacronym{LED}{LED}{light-emitting device}
\newacronym{RIE}{RIE}{reactive ion etching}
\newacronym{ICP}{ICP}{inductively coupled plasma}
\newacronym{IPA}{IPA}{isopropyl alcohol}
\newacronym{BOE}{BOE}{Buffered Oxide Etch}
\newacronym{EBL}{EBL}{electron beam lithography}
\newacronym{RF}{RF}{radio-frequency}
\newacronym{AP}{AP}{atmospheric pressure}
\newacronym{RP}{RP}{reduced-pressure}
\newacronym{OEIC's}{OEIC's}{optoelectronic integrated circuits}
\newacronym{VS}{VS}{virtual substrate}
\newacronym{CW}{CW}{continuous wave}
\newacronym{APD}{APD}{avalanche photodiode}
\newacronym{IoT}{IoT}{internet of things}
\newacronym{PIC}{PIC}{photonic integrated circuit}
\newacronym{CMOS}{CMOS}{complementary metal-oxide-semiconductor}
\newacronym{LSW}{LSW}{\quot{Lasher-Stern-W\"{u}rfel}}

\title{Radiative Carrier Lifetime in \GeSn{1-x}{x} Mid-Infrared Emitters}

\author{G\'erard Daligou}
\affiliation{\polydept{}}

\author{Anis Attiaoui}
\affiliation{\polydept{}}

\author{Simone Assali}
\affiliation{\polydept{}}

\author{Patrick Del Vecchio}
\affiliation{\polydept{}}

\author{Oussama Moutanabbir}
\email{oussama.moutanabbir@polymtl.ca}
\affiliation{\polydept{}}

\begin{abstract}
\medskip
\GeSn{1-x}{x} semiconductors hold the premise for large-scale, monolithic mid-infrared photonics and optoelectronics. However, despite the successful demonstration of several \GeSn{1-x}{x}-based photodetectors and emitters, key fundamental properties of this material system are yet to be fully explored and understood. In particular,  little is known about the role of the material properties in controlling the recombination mechanisms and their consequences on the carrier lifetime. Evaluating the latter is in fact fraught with large uncertainties that are exacerbated by the difficulty to investigate narrow bandgap semiconductors. To alleviate these limitations, herein we demonstrate that the radiative carrier lifetime can be obtained from straightforward excitation power- and temperature-dependent photoluminescence measurements. To this end, a theoretical framework is introduced to simulate the measured spectra by combining the band structure calculations from the \kp{} theory and the \ac{EFA} to estimate the absorption and spontaneous emission. The model computes explicitly the momentum matrix element to estimate the strength of the optical transitions in single bulk materials, unlike the \ac{JDOS} model which assumes a constant matrix element. Based on this model, the temperature-dependent emission from \GeSn{0.83}{0.17} samples at a biaxial compressive strain of $-1.3\%$ was investigated. The simulated spectra reproduce accurately the measured data thereby enabling the evaluation of the steady-state radiative carrier lifetimes, which are found in the $3$-$\SI{22}{\nano\second}$ range for temperatures between $10$ and $\SI{300}{\kelvin}$ at an excitation power of $\SI{0.9}{\kilo\watt\per\square\centi\meter}$. For a lower power of $\SI{0.07}{\kilo\watt\per\square\centi\meter}$, the obtained lifetime has a value of  $\SI{1.9}{\nano\second}$ at $\SI{4}{\kelvin}$. The demonstrated approach yielding the radiative lifetime from simple emission spectra will provide valuable inputs to improve the design and modeling of \GeSn{1-x}{x}-based devices.

\end{abstract}
\maketitle

\section{INTRODUCTION}\label{sec:intro}
\GeSn{1-x}{x} alloys constitute an emerging class of group IV semiconductors providing a tunable narrow bandgap, which has been highly attractive to implement scalable, silicon-compatible mid-infrared photonic and optoelectronic devices \cite{moutanabbirMonolithicInfraredSilicon2021}. This potential becomes increasingly significant with the recent progress in nonequilibrium growth processes enabling high Sn content \GeSn{1-x}{x} layers and heterostructures leading to the demonstration of a variety of monolithic mid-infrared emitters and detectors 
\cite{ bucaRoomTemperatureLasing2022, changMidinfraredResonantLight2022, chretienGeSnLasersCovering2019, chretienRoomTemperatureOptically2022, elbazUltralowthresholdContinuouswavePulsed2020b, joo1DPhotonicCrystal2021, jungOpticallyPumpedLowthreshold2022, li30GHzGeSn2021,atallaAllGroupIV2021b, liuSnContentGradient2022, luoExtendedSWIRPhotodetectionAllGroup2022b, marzbanStrainEngineeredElectrically2022, talamassimolaCMOSCompatibleBiasTunableDualBand2021, tranSiBasedGeSnPhotodetectors2019, xuHighspeedPhotoDetection2019, zhouElectricallyInjectedGeSn2020a,atallaHighBandwidthExtendedSWIRGeSn2022a}. Notwithstanding the recent developments in device engineering, the impact of structural characteristics on the basic behavior of charge carriers is yet to be fully understood. This includes the role of Sn content, lattice strain, and growth defects in shaping the nature and magnitude of the recombination mechanisms and their consequences on the carrier lifetime. Particularly, investigating the latter remains a daunting task due to the lack of methods and tools that can be applied to probe charge carriers in narrow bandgap materials. For instance, time-resolved \ac{PL} can hardly be applied to investigate materials at emission wavelengths in the mid-infrared range as high-speed detectors covering this range are not broadly available. Thus, the very few reported time-resolved studies concern \GeSn{1-x}{x} emitting below $\SI{2.3}{\micro\meter}$ corresponding to a relatively low Sn content and/or highly compressively strained materials \cite{Julsgaard:20,Vitello2020,hudaitHighCarrierLifetimes2022}. 

\par In an attempt to circumvent the aforementioned limitations, a recent study employed time-resolved \ac{PL} with a nonlinear crystal allowing the up-conversion of photons emitted to a shorter wavelength that can be detected by a conventional silicon-based avalanche photodiode \cite{Julsgaard:20}. An effective carrier lifetime of $\SI{217}{\pico\second}$ at $\SI{20}{\kelvin}$ was estimated for \GeSn{0.875}{0.125} with $-0.55\%$ strain using this method \cite{Julsgaard:20}. Additionally, by investigating spin-dependent optical transitions leveraging the Hanle effect under steady-state excitation, systematic studies combining modeling and magneto-\ac{PL} analysis of pseudomorphic layers at a Sn content below $10\%$ reported a radiative lifetime in the $0.5$-$\SI{2.5}{\nano\second}$ range at $\SI{10}{\kelvin}$ \cite{Vitello2020}. However, significantly higher carrier lifetimes reaching $\SI{450}{\nano\second}$ were recently reported for \GeSn{1-x}{x} ($x<0.06$) grown on InAlAs buffer layers as measured by contactless microwave photoconductive decay \cite{hudaitHighCarrierLifetimes2022}. This scarcity of studies on carrier dynamics in narrow bandgap \GeSn{1-x}{x} semiconductors limits the understanding of their fundamental behavior and burdens the development of accurate and predictive models for \GeSn{1-x}{x}-based mid-infrared optoelectronic devices. 

\par In this work, we demonstrate that straightforward \ac{PL} analyses along with the proper theoretical framework are sufficient to alleviate these challenges and extract the radiative carrier lifetime in \GeSn{1-x}{x} mid-infrared emitters and evaluate its evolution as a function of temperature. The approach relies on the simulation of the experimental \ac{PL} spectra by combining the band structure calculations using the \kp{} formalism together with the \ac{EFA} to estimate the absorption and spontaneous emission spectra. Unlike the \ac{JDOS} model, in which the momentum matrix element is considered constant, the oscillator strengths are explicitly computed in this model. 

\par In the following sections, the model is described followed by the experimental demonstration using as-grown \GeSn{0.83}{0.17} layers, emitting at wavelengths above $\SI{3}{\micro\meter}$.

\section{Theoretical framework}
The \ac{PL} spectrum intensity is usually determined using the direct interband emission theory and the spontaneous emission spectrum \cite{saleh2019fundamentals}. Indeed, by considering a slab of homogeneously excited material, Lasher and Stern \cite{LasherStern1964} and W\"{u}rfel \cite{PWurfel1982} expressed the \textit{external} flux of spontaneous radiative emission in terms of the spectral absorptivity under non-equilibrium conditions in terms of the quasi-Fermi level splitting, $\Delta \mu = \mu_e - \mu_h$. The resulting \ac{LSW} equation is
\begin{equation}
    I_{\text{PL}}(E) = \frac{2\pi}{\hbar^3c^2}\frac{E^2a(E)}{\exp(\frac{E - \Delta\mu}{k_{\text{B}}T}) - 1} = \frac{a(E)}{\alpha(E)}\cdot\frac{r^{\text{sp}}(E)}{4n_r^2}
    \label{eqn:PL_LSW}
\end{equation}
where $r^{\text{sp}}$ is the \textit{internal} spontaneous emission spectrum, $\alpha(E)$ is the absorption spectrum and $n_r$ the refractive index of the medium. $a(E)$ is the spectral absorptivity defined as expression \eqref{eqn:LSW_absorptivity} with $R$ the reflection from the outside onto the sample surface, and $d$ the thickness of the conceptual slab \cite{Katahara2016}.
\begin{equation}
    a(E) = (1 - R(E))\Big[1 - \exp\qty(-\alpha(E)d)\Big]
    \label{eqn:LSW_absorptivity}
\end{equation}
Note that $d$ is also considered as a characteristic length scale over which carriers are generated, travel and recombine radiatively \cite{Katahara2016}. Based on this definition, this parameter should be inversely proportional
to the absorption coefficient at the excitation wavelength i.e.\ $d\approx 1/\alpha(\lambda_{\text{laser}})$. However, the \ac{PL} spectrum will mostly be centered around the bandgap energy where the $\alpha(E)$ is at least one or two orders of magnitude smaller than $\alpha(\lambda_{\text{laser}})$, in the case of non-resonant excitation. Therefore $\alpha d<< 1$ and the absorbance $a(E)$ can be simplified by expanding the exponent with a Taylor series such that $a(E)\approx A\alpha(E)$. In that case, the \ac{PL} intensity from equation \eqref{eqn:PL_LSW} becomes
\begin{equation}
    I_{\text{PL}}(E) \approx \frac{2\pi A}{\hbar^3c^2}\frac{E^2\alpha(E)}{\exp(\frac{E - \Delta\mu}{k_{\text{B}}T}) - 1} \approx \frac{A}{4n_r^2}\cdot r^{\text{sp}}(E)
    \label{eqn:PL_LSW_smplified}
\end{equation}
With this approximation, The \ac{PL} spectrum intensity is therefore entirely defined by the internal spontaneous emission spectrum or the absorption spectrum depending on the formula used. 

\par The spontaneous emission spectrum $r^{\text{sp}}$ is calculated using the Fermi's golden rule \cite{chuang2012physics, landsberg_1992} and the perturbation theory as described in equation \eqref{eqn:rsponGeneral}:
\begin{align}
         r^{\text{sp}}(\hbar\omega) & = \frac{\hbar\omega\Lambda}{V}\sum_{c, v, \va*{k}}\abs{\mel**{\Phi_c(\va*{k})}{\frac{\hbar}{m_0}\vu*{\varepsilon}.\va*{p}}{\Phi_v(\va*{k})}}^2f\left(\epsilon_c, \mu_e\right)\nonumber\\
         & \quad \times\delta\left(\epsilon_c(\va*{k}) - \epsilon_v(\va*{k}) - \hbar\omega\right)\left[1-f\qty(\epsilon_v,\mu_h)\right], 
    \label{eqn:rsponGeneral}
\end{align}
where, $\Lambda = n_re^2/\pi c^3\varepsilon_0\hbar^4$ is a material-related constant with $e$ the elementary charge, $n_r$ the refractive index of the material, and $c$ the speed of light in vacuum. $V$ is the volume of the states in the $\va*{k}$-space. The summations are done over the different values of $\va*{k}$ in the \ac{BZ} to account for the possible transitions between the conduction and the valence bands. The Dirac delta distribution is used to limit the transitions to those with an energy difference of $\hbar\omega$, the photon energy. Moreover, $M^2_{i,f}(\va*{k}) = \abs{\mel**{\Phi_f(\va*{k})}{\frac{\hbar}{m_0}\vu*{\varepsilon}.\va*{p}}{\Phi_i(\va*{k})}}^2$ represents the strength of the transition from the state $\ket{\Phi_i}$ to $\ket{\Phi_f}$ with $\vu*{\varepsilon}$ the polarization unit vector and $\va*{p}$ the momentum matrix operator. Finally, the Fermi-Dirac statistic is used to account for the probability occupation of the different states, with $f$ given by $f(\epsilon, \mu) = \left[1 + \exp\left(\frac{\epsilon-\mu}{k_{\text{B}}T}\right)\right]^{-1}$ in which $\epsilon$ is the energy and $\mu$ the Fermi-level of the charge carrier described by the function.

\par The computation of the spontaneous emission spectrum requires prior knowledge of the band structure of the semiconductors, the momentum matrix elements, and the quasi-Fermi levels, as seen in equation \eqref{eqn:rsponGeneral}. In the current literature, for a single bulk direct bandgap semiconductor, $r^{\text{sp}}$ is commonly computed using the \ac{JDOS} model \cite{schubert_2006, Stange2019, Wirths2015}. This model relies on the \ac{PBA}  which leads to a set of relatively easy analytical formulas. It is mostly accurate for a non-degenerately doped semiconductor in weak-injection regime with the quasi-Fermi levels lying within the bandgap and away from the different band edges by several $k_{\text{B}}T$, where $k_B$ is the Boltzmann's constant and $T$ the temperature ($\Delta\mu\approx 0$). This model was extended in \cite{Dijkstra2021} to account for different excitation regimes by explicitly evaluating the quasi-Fermi level splitting $\Delta\mu$, and the non-equilibrium absorption spectrum in equation \eqref{eqn:PL_LSW_smplified}. However, it still relies on the \ac{PBA} which restricts the analysis. Indeed, for higher excitation power and/or doping concentration, $\mu_e$ and $\mu_h$ would shift towards, and even beyond, the band edges where the \ac{PBA} should be less accurate. Besides, for a biaxially strained semiconductor, the $\va*{k}$ direction degeneracies in the \ac{BZ} are expected to be broken. In this situation, the band dispersion would be increasingly anisotropic, thus challenging one of the core principles of the \ac{PBA}. A more accurate theoretical framework is therefore required for the computation of the spontaneous emission spectrum $r^{\text{sp}}$, and the description of the measured \ac{PL} spectra.
\begin{figure}[H]
    \centering
    \includegraphics[width=\columnwidth]{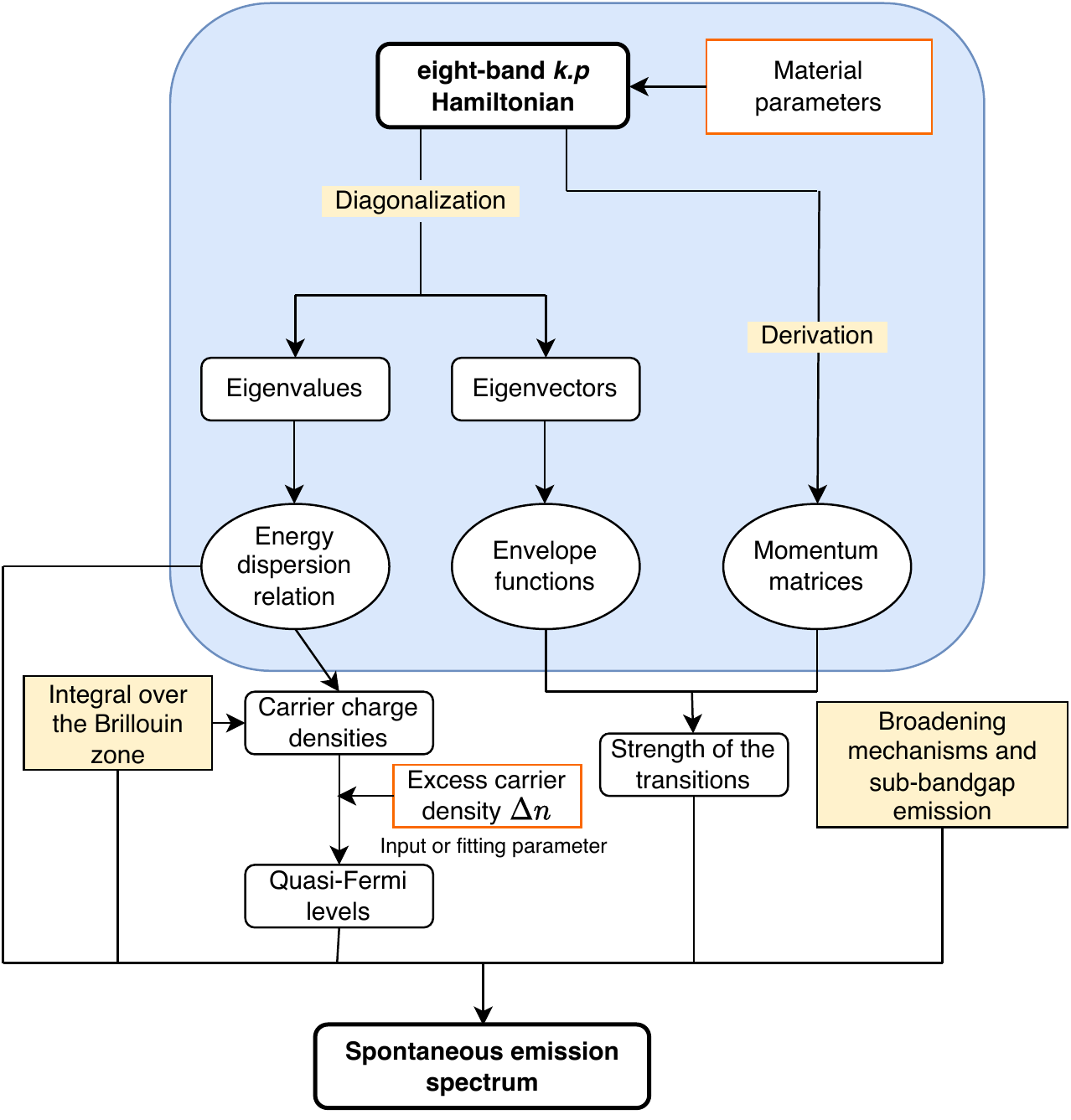}
     \caption{Different steps followed in the computation of the spontaneous emission spectra}
    \label{fig:ComputationMethodology}
\end{figure}
\subsection{Spontaneous emission spectrum and eight-band \kp{} formalism}\label{section:rsp_kp} 
\par The spectrum $r^{\text{sp}}$ is computed using the eight-band \kp{} formalism together with the \ac{EFA} \cite{Bahder1990}, following the simulation workflow summarized in \fref{fig:ComputationMethodology}{}. The eight-band \kp{} \GeSn{1-x}{x} material parametrization is based on early reports \cite{Chang2010,KL2012,Polak_2017}, while strain implementation is based on the Bir-Pikus formalism \cite{PikusBir}. To account for the inaccuracy of the Vegard's law to estimate the bandgaps of \GeSn{1-x}{x} alloys, bandgap bowing parameters are introduced for L and $\Gamma$ high-symmetry points. 
Unlike the \ac{JDOS} model (and all the different models relying on the \ac{PBA}), the evolution of the strength of the optical transitions with the wave vector $\va*{k}$ is explicitly computed using the formalism developed by Szmulowicz \cite{Szmulowicz1995}. If $\ket{\Phi_i}$ and $\ket{\Phi_f}$ are the initial and final states within the \ac{EFA}, the strength of the transition is given by \cite{Szmulowicz1995}:
\begin{align}
        \mel**{\Phi_i}{\frac{\hbar}{m_0}\vu{e}.\va*{p}}{\Phi_f} & = \sum_{\mu,\,\nu}\Phi_{i,\mu}^*(\va*{k})\left(\vu{\varepsilon}\vdot\pdv{\mathcal{H}_{\mu\nu}(\va*{k})}{\va*{k}}\right)\Phi_{f,\nu}(\va*{k}) \nonumber\\ 
        & = \sum_{\mu,\,\nu,\,l}\Phi_{i,\mu}^*(\va*{k})\left[\varepsilon_l\cdot\left(\pdv{\mathcal{H}_{\mu\nu}}{k_l}\right)\right]\Phi_{f,\nu}(\va*{k}),
    \label{eqn:MomentumMatrix1}
\end{align}
where $\Phi_{i,\mu}$ and $\Phi_{f, \nu}$ are the coefficients of the envelope function vector related to the states $\ket{\Phi_i}$ and $\ket{\Phi_f}$, respectively. The unit vector $\vu{\varepsilon}$ gives the polarization of the incident light, while $\partial\mathcal{H}_{\mu\nu}(\va*{k})/\partial\va*{k}$ is the derivative of the \kp{} Hamiltonian with respect to the wave vector $\va*{k}$. The expressions of the different momentum matrices $\partial\mathcal{H}_{\mu\nu}/\partial k_l$ can be found in the \mentionSupp{1} \citesupp.

\par For a given value of the optically injected carrier density $\Delta n$, if $n_0$ and $p_0$ denote the total electrons and holes' charge densities at thermal equilibrium, the quasi-Fermi levels $\mu_e$ (for electrons) and $\mu_h$ (for holes) are determined by solving the set of equations \eqref{eqn:BulkQuasiFermi}:
\begin{align}
    \begin{aligned}
        n_0 + \Delta n &= \frac{1}{(2\pi)^3}\sum_{i \in \mathrm{CB}}\int_{\text{BZ}} \frac{d^3\va*{k}}{1 + \exp\left(\frac{\epsilon_i(\va*{k}) - \mu_e}{kT}\right)}\\
        p_0 + \Delta n &= \frac{1}{(2\pi)^3}\sum_{i \in \mathrm{VB}}\int_{\text{BZ}} \frac{d^3\va*{k}}{1 + \exp\left(\frac{\mu_h - \epsilon_j(\va*{k})}{kT}\right)} 
    \end{aligned}   
    \label{eqn:BulkQuasiFermi}
\end{align}
Herein, the conduction band electrons are assumed to be shared between the $\Gamma$ and $L$ valleys. This assumption is only relevant when the energy band offset between these valleys is relatively close to the thermal energy $k_{\text{B}}T$ to enable the electrons to transition between them. The carrier concentration $n_0$ and $p_0$ are evaluated after solving the electroneutrality equation to estimate the thermal equilibrium Fermi level $E_f$. Besides, the computation of the integrals over the \ac{BZ}, required for estimating the quasi-Fermi levels and $r^{\text{sp}}$, relies on the \ac{SLA} \cite{Enders1996a}. Within this approximation, the three-dimensional \ac{BZ} integrals are replaced by a sum of one-dimensional integrals over some characteristic directions (denoted as \quot{special}) of the crystal lattice. These directions could, for example, be the symmetry directions used in the eight-band \kp{} formalism. If we denote by $\mathcal{L}$ the set of the special directions, the electrons density from equation \eqref{eqn:BulkQuasiFermi} becomes 
\begin{align}
    n &= \frac{1}{2\pi^2}\sum_{\substack{D \in \mathcal{L}\\i \in \mathrm{CB}}}w_{D}\left( \int_0^{k_{\text{BZ}}} \frac{k^2_{D}\dd k_{D}}{1 + \exp\left(\frac{\epsilon_i(k_{D}) - \mu_e}{kT}\right)}\right)
    \label{eqn:SpecialLineDensities}
\end{align}
with $w_{D}$ the weight of the direction $D$, $\epsilon_i(k_{\text{D}})$ the energy of the conduction band $i$ at $k_{\text{D}}$ and $k_{\mathrm{BZ}} \sim 0.5$ (units of $\pi/a_0$, $a_0$ being the lattice constant of the material) for the eight-band $\kp$ model to still be accurate. Depending on the computation, the exact value of the upper limit $k_{\text{BZ}}$ could be neglected since the integrands are expected to vanish rapidly while increasing the value of $k_{\text{D}}$. More information about all the different directions considered in our framework can be found in the \mentionSupp{2} \cite{Note1}. Unlike the \ac{PBA}, which leads to parabolic and isotropic-like band structure, this method accounts for the anisotropy and the non-parabolicity of the bands obtained with the \kp{} theory and approximates the warping of real bands. It is, therefore, expected to be more accurate. 
\par The theoretical spontaneous emission spectrum developed previously is often insufficient to accurately describe the \ac{PL} spectrum. Indeed, the sub-bandgap emission resulting from carrier disorders and broadening mechanisms in the materials is not accounted for in equation \eqref{eqn:rsponGeneral} \cite{PhysRevB.30.813, PhysRevB.41.12096, Asada1989}. To include these contributions, the theoretical spectrum $r^{\text{sp}}_{\text{ideal}}$ from equation \eqref{eqn:rsponGeneral} is convoluted with a broadening function $\mathcal{B}$, as outlined in equation \eqref{eqn:Rsp_Broadening}. 
\begin{equation}
    r^{\text{sp}}(\hbar\omega) =  \int_{-\infty}^{+\infty}r^{\text{sp}}_{\text{ideal}}(\epsilon)\cdot\mathcal{B}(\hbar\omega - \epsilon)\mathrm{d}\epsilon
    \label{eqn:Rsp_Broadening}
\end{equation}
The broadening function $\mathcal{B}$ is usually chosen as a Gaussian or a Lorentzian to account for the inhomogeneous and homogeneous broadening mechanisms, respectively. However, the Lorentzian function was reported to sometimes overestimate the effects of the homogeneous broadening due to its slowly decaying tails. For that reason, it is usually replaced by a hyperbolic secant function \cite{chowSemiconductorlaserPhysics1997}.
\begin{figure}[H]
    \centering
    \includegraphics[width=\columnwidth]{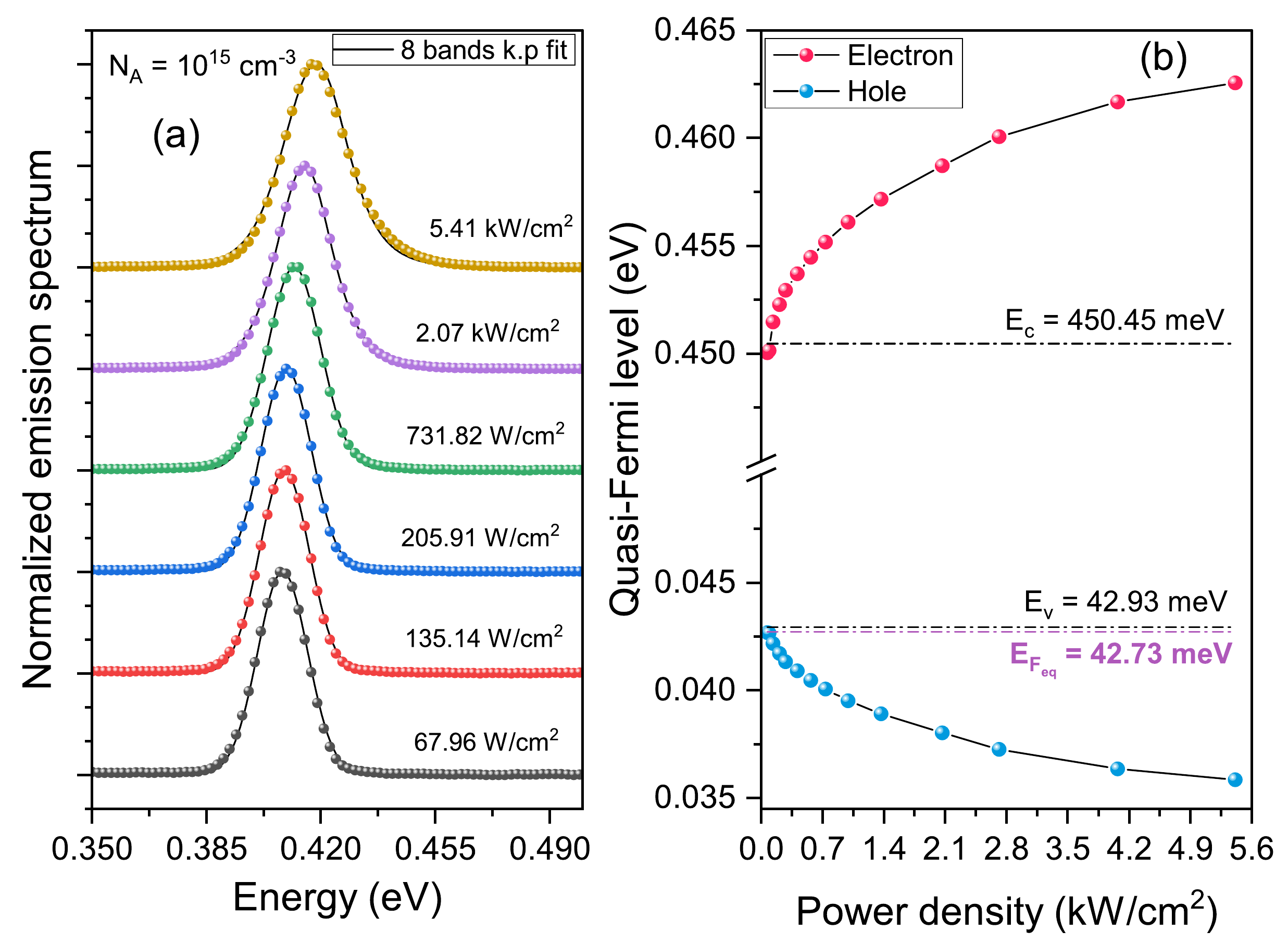}
    \caption{(a) Power-dependent photoluminescence at $\SI{4}{\kelvin}$ for the -1.27\% biaxially strained \GeSn{0.83}{0.17}. The scatter points are from the measurements while the black lines are the results from the simulations. (b) Evolution of the extracted quasi-Fermi levels ($\mu_e,\, \mu_h$) with the power density.}
    \label{fig:Power-PL-comparison-4K-1}
\end{figure}
\subsection{Spontaneous emission intensity and steady-state radiative carrier lifetime}
We shall use the symbol $R_{\text{sp}}^{\vu{\varepsilon}}$ to denote the rate of polarization-dependent spontaneous emission per unit volume, where $\vu{\varepsilon}$ gives the polarization of the incident light. The quantity $R_{\text{sp}}^{\vu{e}}$ is generally defined as the integral of the polarization-dependent spontaneous emission spectrum over the range of photon energy bigger than the band-gap of the material, and given by equation \eqref{eqn:SponRate_Polarization} \cite{chuang2012physics}.
\begin{equation}
    R_{\text{sp}}^{\vu{\varepsilon}} = \int_0^{+\infty}r^{\text{sp}}_{\vu{\varepsilon}}(\hbar\omega)\dd{\hbar\omega}
     \label{eqn:SponRate_Polarization}
\end{equation}
For an unpolarized incident light, the total spontaneous emission rate per unit volume $R_{sp}$ is defined as the average of the contributions from the three polarizations defined by the unit vectors $\vu{\varepsilon}_x = (1,0,0)$, $\vu{\varepsilon}_y$ and $\vu{\varepsilon}_z$. 
\par The steady-state radiative carrier lifetime $\tau_{\text{rad}}$ is determined by the net rate of spontaneous emission $R_{\text{sp}}^{\text{net}}$ and the density of photo-excited carriers $\Delta n$ (equation \eqref{eqn:tauRad}). $R_{\text{sp}}^{\text{net}}$ is defined as the amount by which the non-equilibrium spontaneous recombination rate $R_{\text{sp}}$ exceeds the thermal equilibrium generation rate $G_0$, which is the same as the thermal equilibrium spontaneous emission rate. 
\begin{align}
    \tau_{\text{rad}} &= \frac{\Delta n}{R_{\text{sp}}^{\text{net}}} 
     \label{eqn:tauRad}
\end{align}
$R_{\text{sp}}^{\text{net}}$ is usually estimated using equation \eqref{eqn:ApproxRsp}, in which $B$ is a material-dependent parameter known as the bimolecular recombination coefficient. In that case, $\tau_{\text{rad}}$ becomes relatively easy to compute, as presented in equation \eqref{eqn:tauRad_Approx}.
\begin{equation}
    R_{\text{sp}}^{\text{net}} = R_{\text{sp}}^{\text{neq}} - R_{\text{sp}}^{\text{eq}} = B(np - n_0p_0)
    \label{eqn:ApproxRsp}
\end{equation}
\begin{align}
    \tau_{\text{rad}} &= \frac{1}{B\qty(\Delta n + n_0 + p_0)} 
     \label{eqn:tauRad_Approx}
\end{align}
The bimolecular recombination coefficient $B$ is typically assumed to be independent of $\Delta n$ (and, therefore, the quasi-Fermi levels). However, this approximation is not always accurate. For example, $B$ was previously shown to vary linearly with the excess carrier density $\Delta n$ in III-V semiconductors \cite{sternCalculatedSpectralDependence1976, suCarrierDependenceRadiative1984,Olshansky1984}. For that reason, it is reasonable to rely only on equations \eqref{eqn:SponRate_Polarization} and \eqref{eqn:tauRad}, which state the general case without any specific approximations.

\section{Results and discussion}\label{sec:resultsDiscussion}
The accuracy of the established theoretical framework has been evaluated through the analysis of the \ac{PL} spectra recorded as a function of the excitation power and temperature from \GeSn{0.83}{0.17} layers \cite{Assali2021}. The epitaxial growth of these layers was achieved using \ac{LP-CVD} starting from a $600$ - $\SI{700}{\nano\meter}$ Ge virtual substrate on a $\SI{4}{inch}$ Si wafer. To ensure the growth of a \GeSn{0.83}{0.17} layer with a uniform Sn composition, a multilayer heterostructure consisting of \ac{TL}/\ac{ML}/\ac{BL} was grown while the incorporation of Sn in each layer is controlled by adjusting the growth temperature.  More details on the growth and characterization of \GeSn{0.83}{0.17} material can be found in \cite{Assali2021}.

\par In as-grown \GeSn{0.83}{0.17} layers, the band alignment favors the electrons and holes diffusion to the \ac{TL}, where they should recombine. Indeed, the \ac{PL} spectra are confirmed to originate from carrier recombination in this specific layer \cite{Assali2021}. Therefore, from a theoretical standpoint, it would be judicious to analyze the \ac{PL} results as if they were emitted by a bulk \GeSn{}{} material with a 17 at.\ \% Sn composition.  On this basis, the different power-dependent \ac{PL} spectra, recorded at 4 K, were simulated by iteratively evaluating $r^{\text{sp}}$ as well as the excess carrier concentration $\Delta n$ and $\gamma$ the \ac{FWHM} of the broadening function. To solve equations \eqref{eqn:BulkQuasiFermi}, and extract the quasi-Fermi levels couple ($\mu_e,\,\mu_h$), the p-type background doping was considered to be around $\SI[per-mode=reciprocal]{e15}{\per\cubic\centi\meter}$ at $\SI{4}{\kelvin}$. This value was chosen with reference to the p-type background doping estimated between $\SI{1e17}{}$ and $\SI[per-mode=reciprocal]{5e17}{\per\cubic\centi\meter}$ at 300 K \cite{atallaAllGroupIV2021}. 
\begin{figure}[H]
    \centering
    \includegraphics[width=\columnwidth]{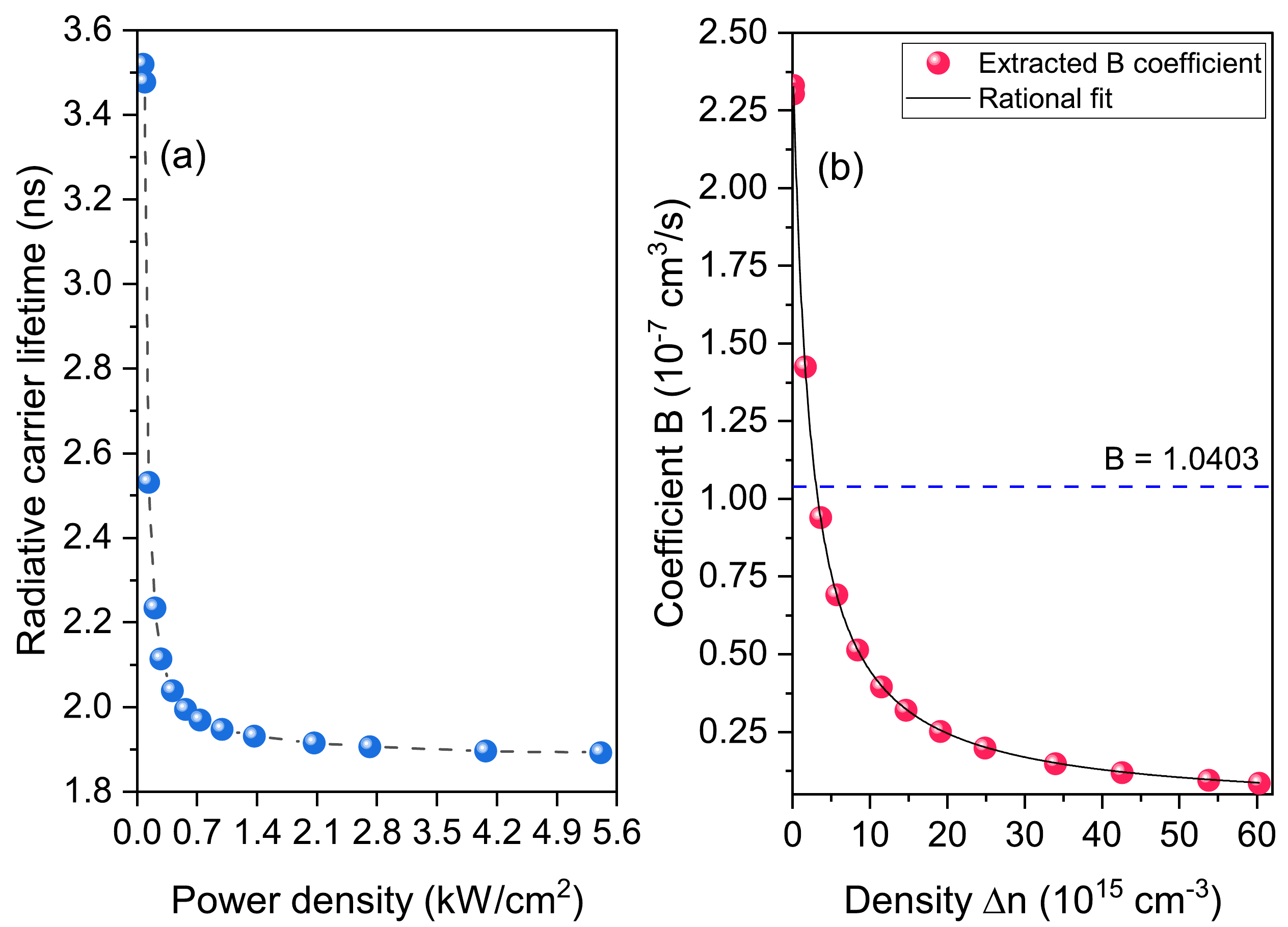}
    \caption{(a) Evolution of the radiative carrier lifetime as a function of the excitation power density $P_{\text{exc}}$, (b) Evolution of the $B$ coefficient as a function of the excess carrier density $\Delta n$. The solid spheres represent the values of $B$ computed from the extracted $\Delta n$, $R_{\text{sp}}$ using equation \eqref{eqn:ApproxRsp}, while the black line is the result of a fit using a rational function.}
    \label{fig:Power-PL-comparison-4K-2}
\end{figure}
\par \fref{fig:Power-PL-comparison-4K-1}{(a)} displays the measured and simulated spectra for the as-grown \GeSn{0.83}{0.17} material. For each power density, a coefficient of determination ($R^2$) of around $99.5\%$ is obtained, thus highlighting the accuracy of the simulated spectra. Moreover, the evolution of the extracted quasi-Fermi levels with the excitation power density $P_{\text{exc}}$ is outlined in \fref{fig:Power-PL-comparison-4K-1}{(b)}.
\begin{figure}[htb]
    \centering
    \includegraphics[width=\columnwidth]{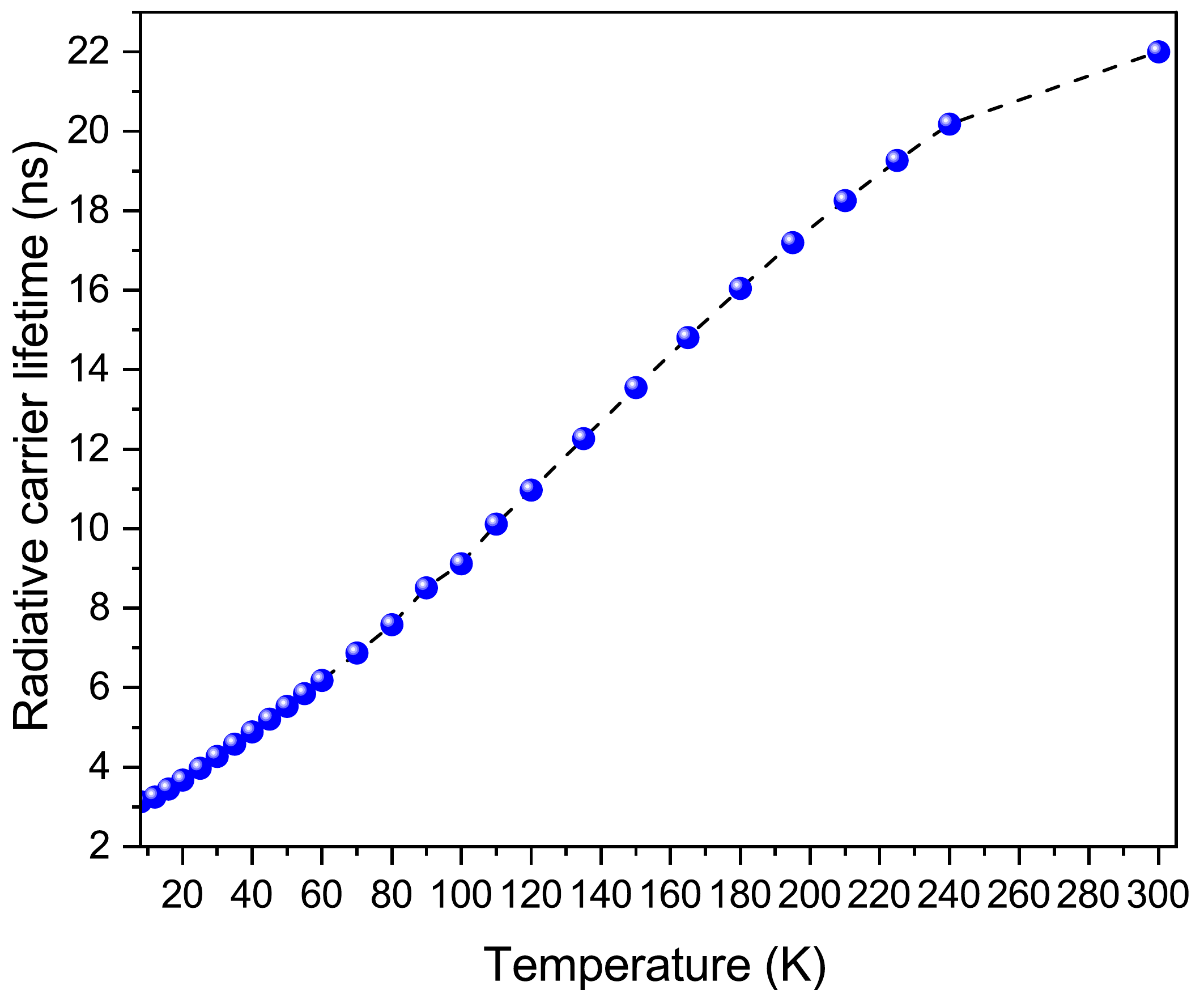}
    \caption{Evolution of the radiative carrier lifetime as function of temperature for the as-grown \GeSn{0.83}{0.17}.}
    \label{fig:TD-PL-fig2}
\end{figure}
For a p-type background doping of $\SI[per-mode=reciprocal]{e15}{\per\cubic\centi\meter}$, the thermal equilibrium Fermi level $E_{\text{F}}$ is about $\SI{42.73}{\milli\electronvolt}$. As shown in \fref{fig:Power-PL-comparison-4K-1}{(b)}, the non-degenerate semiconductor approximation is not appropriate here since $E_{\text{F}}$ is less than the top valence band edge located around $\SI{42.93}{\milli\electronvolt}$. Starting from a power density of $\SI{67.95}{\watt\per\square\centi\meter}$, both the electrons and holes quasi-Fermi levels start to deviate from $E_{\text{F}}$. In fact, a progressive increase from $450.02$ to $\SI{462.57}{\milli\electronvolt}$ is observed for the quasi-Fermi level $\mu_e$, causing the electron concentration to increase. Simultaneously, the holes quasi-Fermi level $\mu_h$ decreases while remaining very close to the thermal equilibrium level with a maximum offset of $\SI{6.88}{\milli\electronvolt}$ at $\SI{5.4}{\kilo\watt\per\square\centi\meter}$. While these variations may be perceived as small, they are not insignificant. Indeed, with the thermal energy of about $\SI{0.34}{\milli\electronvolt}$ at 4 K, one should expect a noticeable increase in the spontaneous emission intensity $R_{\text{sp}}$. Moreover, using equation \eqref{eqn:tauRad}, the steady-state radiative carrier lifetime $\tau_{\text{rad}}$ was extracted (\fref{fig:Power-PL-comparison-4K-2}{(a)}) and shown to decrease from $\SI{3.52}{}$ to $\SI{1.89}{\nano\second}$ in the range of power density used in this study.

\par Besides, the accuracy of equation \eqref{eqn:ApproxRsp} was also evaluated using the different parameters obtained from the analysis above. Indeed, the bimolecular recombination coefficient $B$ was computed from the extracted values of $R_{sp}$ and $\Delta n$ and its behavior is outlined in \fref{fig:Power-PL-comparison-4K-2}{(b)}. Rather than being constant, it decreases with $\Delta n$, as suggested earlier for III-V semiconductors \cite{sternCalculatedSpectralDependence1976, suCarrierDependenceRadiative1984}. However, its evolution for the as-grown \GeSn{0.83}{0.17} is not as linear as presented by Olshansky et al.\ for InGaAsP and AlGaAs light sources \cite{Olshansky1984}. In fact, after performing a fit of the obtained data, $B$ was shown to evolve with $\Delta n$ following a rational function (\fref{fig:Power-PL-comparison-4K-2}{(b)}). Additionally, for $\Delta n$ above $\SI[per-mode=reciprocal]{e15}{\per\cubic\centi\meter}$, the values extracted were shown to be lower than the value of $\SI{1.04e-7}{\cubic\centi\meter\per\second}$ computed assuming parabolic band dispersion and the non-degenerate semiconductor approximation.

\par The impact of temperature on the steady-state radiative carrier lifetime has also been investigated. Herein, assuming a $p$-type background doping of $\SI[per-mode=reciprocal]{1e15}{\per\cubic\centi\meter}$ at $\SI{4}{\kelvin}$ and $\SI[per-mode=reciprocal]{1e17}{\per\cubic\centi\meter}$ at $\SI{300}{\kelvin}$, which is in line with recent measurements \cite{atallaAllGroupIV2021}, the evolution of the doping with temperature was estimated. Using these values, the temperature-dependent \ac{PL} spectra were simulated with the theoretical estimation of the spontaneous emission spectrum from the framework described above, and the evolution of $\tau_{\text{rad}}$ was extracted for the as-grown \GeSn{0.83}{0.17}, as displayed in \fref{fig:TD-PL-fig2}{}. Note that from this analysis, a minimum $R^2$ factor of about $98\%$ was observed throughout the $4$-$\SI{300}{\kelvin}$ range. The estimated steady-state radiative carrier lifetime $\tau_{\text{rad}}$ increases with the temperature from $\sim\SI{3.2}{\nano\second}$ at $\SI{10}{\kelvin}$ to $\sim\SI{22.2}{\nano\second}$ at $\SI{300}{\kelvin}$. These values are very comparable to the reported recombination lifetimes in literature for III-V compound semiconductors, which are generally in the nanoseconds range \cite{Feldmann1987, Bellesa1998, Hooft1985}. They are also of the same order of magnitude as the values for Ge calculated from first principles \cite{Rodl2019}. Indeed, the radiative lifetime for Ge in the diamond structure was shown to be around $\SI{10}{\nano\second}$ for $T$ below $\SI{300}{\kelvin}$.
\par Finally, to appreciate the radiative emission strength of \GeSn{1-x}{x} with respect to other direct bandgap semiconductors, we compare the radiative emission rate or more precisely the bimolecular recombination coefficient $B$. Using the same process as for the power-dependent \ac{PL}, $B$ is extracted as a function of the temperature. From this analysis, $B$ is found to evolve following the allometric power law $aT^b$ with $b\approx -1.5143$, and reaching $\SI{3.81e-10}{\cubic\centi\meter\per\second}$ at $\SI{240}{\kelvin}$. This value is comparable to those extracted at $\SI{300}{\kelvin}$ for GaAs ($\SI{3.5e-10}{\cubic\centi\meter\per\second}$), InP ($\SI{1.2e-10}{\cubic\centi\meter\per\second}$), and hexagonal Si$_{0.20}$Ge$_{0.80}$ ($\SI{0.7e-10}{}-\SI{11e-10}{\cubic\centi\meter\per\second}$).


\section{CONCLUSION}\label{sec:concl}
    
To circumvent the limitations in the experimental studies of carrier dynamics in narrow bandgap \GeSn{1-x}{x} materials, this work demonstrates a straightforward method to obtain the carrier radiative lifetime from simple PL spectra. The approach relies on a theoretical framework combining the band structure calculations using the \kp{} formalism together with the envelope function approximation to estimate the absorption and
spontaneous emission spectra. This framework simulates accurately the experimental measurements thereby allowing the evaluation of the steady-state radiative carrier lifetime from the net rate of spontaneous emission and the density of photo-excited carriers. For a \GeSn{0.83}{0.17} material under an in-plane biaxial compressive strain $\varepsilon_{\parallel} = -1.3\%$, the analysis revealed a lifetime $\tau_{\text{rad}}$ in the nanoseconds range increasing from 3 to $\SI{22}{\nano\second}$ for temperatures between 10 and $\SI{300}{\kelvin}$. Additionally, the introduced model also solves the restrictions that are inherent to the joint density of states (JDOS) model resulting from the parabolic band approximation (\ac{PBA}) and the weak-injection approximation.


\paperSection{ACKNOWLEDGEMENTS}
O.M. acknowledges support from NSERC Canada (Discovery, SPG, and CRD Grants), Canada Research Chairs, Canada Foundation for Innovation, Mitacs, PRIMA Qu\'ebec, Defence Canada (Innovation for Defence Excellence and Security, IDEaS), the European Union’s Horizon Europe research and innovation programme under grant agreement No 101070700 (MIRAQLS), and the US Army Research Office Grant No. W911NF-22-1-0277.

\paperSection{AUTHORS INFORMATION}
Corresponding Author:\\
\textcolor{blue}{$^{\dagger}$} \href{mailto:oussama.moutanabbir@polymtl.ca}{oussama.moutanabbir@polymtl.ca}\\
Notes:\\
The authors declare no competing financial interest.

\bigskip

\bibliography{main.bbl} 
\bibliographystyle{apsrev4-2} 

\end{document}